\documentclass[twocolumn]{IEEEtran}

\usepackage{amsthm,amssymb,graphicx,graphicx,multirow,cite,url,color}
\usepackage[cmex10]{amsmath}

\newtheorem{defn6}{definition} \newtheorem{definition}[defn6]{Definition}
\newtheorem{defn2}{theorem} \newtheorem{theorem}[defn2]{Theorem}
\newtheorem{defn5}{corollary} \newtheorem{corollary}[defn5]{Corollary}
\theoremstyle{definition}

\def\ch#1{{#1}}
\def\chr#1{{#1}}

\newcommand{\ceil}[1]{\lceil #1\rceil}

\def\figref#1{Fig.\,\ref{#1}}%
\def\E{\mathbb{E}^{!o}}

\def\P{\mathbb{P}}
\def\p{\mathsf{P}}
\def\ie{{\em i.e.}}
\def\eg{{\em e.g.}}
\def\VV{\mathbf{V}}

\def\d{\mathrm{d}}

\def\R{\mathbb{R}}

\def\Z{\mathbb{Z}}

\def\sinr{\mathsf{SINR}}

\def\i{\mathbf{1}}

\def\x{\mathsf{x}}
\def\S{\mathsf{S}}
\def\I{\mathtt{I}}

\def\y{y}

\def\lt{\lambda_\mathrm{t}}

\def\EP{\mathbb{E}^{!o}}
\def\l{\ell}

\def\T{\theta}
\def\G{\Phi_\eta} 

  \newcommand{\h}[1]{\ensuremath{\mathsf{h}_{#1}}}

\usepackage{color,cite,subfigure}
\begin{document}

\title{Outage Probability of General Ad Hoc Networks\\in the High-Reliability Regime}

\author{Riccardo Giacomelli, Radha Krishna Ganti, \IEEEmembership{Member, IEEE}, and Martin Haenggi, \IEEEmembership{Senior
Member, IEEE}
\thanks{R. Giacomelli is with the Politecnico di Torino, Italy, R. K. Ganti is with the University of Texas at Austin, TX, USA, and M. Haenggi
is with the University of Notre Dame, IN, USA. The contact author is M. Haenggi, {\tt mhaenggi@nd.edu}. This work has been supported
by the NSF (grants CNS 04-47869, CCF 728763) and the DARPA/IPTO IT-MANET program
(grant W911NF-07-1-0028). Manuscript date: \today.}}


\maketitle

\begin{abstract}
Outage probabilities in wireless networks depend on various factors: the node distribution,
the MAC scheme, and the models for path loss, fading and transmission
success. In prior work on outage characterization for networks with randomly placed nodes,
most of the emphasis was put on networks whose nodes
are Poisson distributed and where ALOHA is used as the MAC protocol.
In this paper, we provide a general
framework for the analysis of outage probabilities in the high-reliability regime.
The outage probability characterization is based on two parameters:
the intrinsic {\em spatial contention} $\gamma$ of the network,
introduced in \cite{net:Haenggi09twc}, and the coordination level achieved by the
MAC as measured by the {\em interference scaling exponent} $\kappa$ introduced in this paper.
We study outage probabilities under the
signal-to-interference ratio (SIR) model, Rayleigh fading, and power-law path loss,
and explain how the two parameters depend on the network model. The main result is that
the outage probability approaches $\gamma\eta^{\kappa}$ as the density of interferers
$\eta$ goes to zero, and that $\kappa$ assumes values in 
the range $1\leq \kappa\leq \alpha/2$ for all practical MAC protocols, where $\alpha$ is the path loss exponent.
This asymptotic expression is valid for all motion-invariant point processes.
We suggest a novel and complete taxonomy of MAC protocols based mainly on the value of $\kappa$.
Finally, our findings suggest a conjecture that  bounds the outage probability for all
interferer densities.
\end{abstract}

\section{Introduction}
The outage probability is the natural metric for large wireless systems, where it cannot be assumed that
the transmitters are aware of the states of all the random processes governing the system and, consequently,
nodes cannot adjust their transmission parameters to achieve fully reliable communication. In many networks,
the node locations are a main source of uncertainty, and thus they are best modeled using a stochastic point
process model whose points represent the locations of the nodes. 

Previous work on outage characterization in networks with randomly placed nodes has mainly focused on the
case of the homogeneous Poisson point process (PPP) with ALOHA
(see, \eg, \cite{net:Baccelli06,net:Weber05tit,net:Haenggi09jsac}),
for which a simple closed-form
expression for the outage exists for Rayleigh fading channels.
Extensions  to models with dependence (node repulsion or attraction) are non-trivial. On the repulsion or hard-core side,
where nodes have a guaranteed minimum distance, approximate expressions were derived
in \cite{net:Busson09inria,net:Haenggi08now,net:Baccelli09now2}; on the attraction or clustered side, \cite{net:Ganti09tit} gives an outage
expression in the form of a multiple integral for the case of Poisson cluster processes. 

Clearly, outage expressions for general networks and MAC schemes would be highly desirable. However, the
set of transmitting nodes is only a Poisson point process if all nodes form a PPP {\em and} ALOHA is used.
In all other cases, including, \eg, CSMA on a PPP or ALOHA on a cluster process, the transmitting set is not
Poisson and, in view of the difficulties of analyzing
non-Poisson point processes, it cannot be expected that general closed-form expressions exist. In this paper,
we study outage in general motion-invariant (stationary and isotropic) networks by resorting to
the asymptotic regime, letting the density of interferers $\eta$ go to zero. We will show that
the outage probability approaches $\gamma \eta^\kappa$ as $\eta\rightarrow 0$, where $\gamma$ is the
network's {\em spatial contention} parameter \cite{net:Haenggi09twc}, and $\kappa$ is the {\em interference scaling exponent}. 
\ch{The spatial contention parameter quantifies the network's capability of spatial reuse. It depends on the geometry of
concurrent transmitters, but not on their intensity. The interference scaling exponent, on the other hand, captures how much
the intensity of transmitters affects the outage probability.}
\ch{Denoting by $\p_\eta$ the success probability of the typical link and letting
$\p_0\triangleq \lim_{\eta\to 0} \p_\eta$, the two parameters are formally defined as
follows:}
\begin{definition}[\ch{Interference scaling exponent $\kappa$}]
\ch{
The interference scaling exponent is
\[ \kappa\triangleq\lim_{\eta\to 0} \frac{\log(|\p_0-\p_\eta|)}{\log \eta} \]}
\end{definition}
\begin{definition}[\ch{Spatial contention parameter $\gamma$}]
\ch{The spatial contention is
\[ \gamma\triangleq\lim_{\eta\to 0}\frac{\p_0-\p_\eta}{\eta^\kappa} \,.\]}
\end{definition}

\ch{Note that in most cases $\p_0=1$.}
Interestingly,
$\kappa$ is confined to the range $1\leq \kappa\leq\alpha/2$ for any practical MAC scheme.
While $\kappa=1$ is the exponent for ALOHA, $\kappa=\alpha/2$ can be achieved with MAC schemes that impose a hard
minimum distance between interferers that grows as $\eta$ decreases.


We adopt the standard signal-to-interference-plus noise (SINR)
model for link outages (aka the physical model), where a transmission is
successful if the instantaneous SINR exceeds a threshold $\theta$. With Rayleigh fading,  the success probability is
known to factorize into a term that only depends on the noise and a term that only depends on the interference
\cite{net:Zorzi95,net:Haenggi05twc,net:Baccelli06}:
\begin{align*}
   \P(\sinr>\theta)&=\P(\S>\theta(I+W))\\
    &=\exp(-\theta W/P)\underbrace{\mathbb{E}\exp(-\theta I/P)}_{\chr{\p_\eta}}\,, 
 \end{align*}
where $P$ is the transmit power, $\S$ the received signal power, assumed exponential with mean \ch{$P$} (unit link distance),
$W$ the noise power, and $I$ the interference (the sum of the powers of all non-desired transmitters).
The first term is the noise term, the second one, \ch{denoted as $\p_\eta$}, is the Laplace transform of the
interference, which does not depend on $W$ or $P$.  \ch{$\p_\eta$ is not affected by the
transmit power $P$, since both interference $I$ and desired signal strength $S$ scale with $P$, and their ratio,
the SIR, is independent of $P$.} Since the first term is a pure point-to-point term that does
not depend on the interference or MAC scheme 
we will focus on the second term. By ``high reliability" we mean that \chr{$\p_\eta\approx 1$}, keeping in mind that the
total success probability may be smaller due to the noise term, which can be made arbitrarily close to $1$ \ch{by
choosing a high transmit power}.

\ch{Note that the fading model is a block fading model, \ie, the SINR is not averaged over the fading process. This is
justified in all cases except when nodes are highly mobile, data rates are low, packets are long, {\em and} wavelengths
are short.}

The rest of the paper is organized as follows: In Section \ref{sec:Assumptions}, we introduce the system model.
Section \ref{sec:outage_prob} is the main analytical section, consisting of 3 theorems; the first theorem states
the fundamental bounds on the interference scaling parameter $\kappa$, while the other two show how the
lower and upper bounds can be achieved. Section \ref{sec:examples} presents examples, simulation
results, \ch{and several extensions to the model}, and Section \ref{sec:conclusions} presents the proposed taxonomy of MAC schemes and conclusions,
including a conjecture that provides general upper and lower bounds on the outage probability for all densities
of interferers.

\section{System Model}
\label{sec:Assumptions}
The nodes locations are modeled as a motion-invariant point process 
$\Phi$ of density $\lambda$  on the plane \cite{net:Stoyan95,net:Daley-VereJones07,net:Kallenberg86}.
We assume that the time is slotted,  and that at every time instant a subset of
these nodes $\G$, selected by the MAC protocol, transmit. We
constrain the MAC protocols to have the following properties:
\begin{itemize}
  \item The MAC protocol has some tuning parameter $0\leq\eta\leq 1$ (for example the probability
	of transmission in ALOHA) so that the density of transmitters $\lt$ can be varied from
	$0$ to $\lambda$.
  \item At every time instant the transmitting set $\G\subset \Phi$ is itself a motion-invariant
	 point process  of density
	$\lt=\eta\lambda$.
\end{itemize}
\chr{The transmitter set being a motion-invariant process is not a restrictive condition. In fact, any MAC protocol that is decentralized, fair and location-unaware results in a transmitter set that is motion-invariant.}
The ratio $\eta\triangleq \lt/\lambda$ denotes the fraction of nodes that transmit. \chr{ Table \ref{tab:eta} (at the end of Section
\ref{sec:examples}) illustrates the values of $\eta$ for different MAC protocols. }
 The path-loss model, denoted by 
$\l(x):\ \R^{2}\setminus\{o\}\rightarrow\R^{+}$, is a continuous, positive, non-increasing function of $\Vert x\Vert$   and 
\begin{equation}
\int_{\R^{2}\setminus B(o,\epsilon)}\l(x)\d x<\infty,\quad \forall\epsilon>0\,,
\label{pathloss_cond}
\end{equation}
where $B(o,r)$ denotes the ball of radius $r$ around the origin $o$.
We assume $\l(x)$  to be a power law in one of the forms:
\begin{enumerate}
  \item Singular path loss model: $\Vert x\Vert^{-\alpha}$.
  \item Bounded (non-singular) path loss model: $(1+\Vert x\Vert^{\alpha})^{-1}$ or $\min\{1,\Vert x\Vert^{-\alpha}\}$.
\end{enumerate}
To satisfy the condition \eqref{pathloss_cond}, we require $\alpha>2$ in all the
above models.

Next, to specify the transmitter-receiver pair under consideration,
select a node $\y\in\G$ and let it be the receiver of a virtual
transmitter $z$ at a distance such that $\l(\y-z)=1$. Including the receiver $\y$ as part of
the process $\G$ allows to study the  success probability  at the
receiver rather than at the transmitter and accounts for the spacing of the
transmitters. The success probability obtained is a good approximation for
{\em transmitter-initiated} MACs if $\left\|\y-z\right\|$ is small, since the interference power
level at the receiver is approximately the same as the one at the
transmitter if $\lt^{-1/2}\gg 1$, which certainly holds for small $\eta$.
The analysis in the subsequent sections does not change significantly if the
positions of the transmitter and the receiver are interchanged \ch{(see Section \ref{sec:extensions})}.
Furthermore, the transmission powers at all nodes are chosen to be identical,
to isolate the effect of $\eta$ on the success probability. 
Let $\S$ be the received power from the intended transmitter;  since the fading is Rayleigh,
$\S$ is exponentially distributed with unit mean. Let  $\I(y)$  denote the
interference  at the receiver 
\begin{equation} \label{eq:interf}
\I(y)=\sum_{\x\in\G}\h{\x}\l(\left\|\x-y\right\|),
\end{equation}
where $\h{\x}$ is iid exponential fading with unit mean. Without loss of generality, we can assume that the
virtual receiver is located at $y=0$ and hence 
the probability of success is given by 
\begin{equation} \label{eq:ps}
  \p_\eta\triangleq \mathbb{P}^{!o}\left(\frac{\S}{\I(o)}\geq \theta\right),\quad \theta>0,
\end{equation}
where $\P^{!o}$ is the reduced Palm probability of $\G$. The Palm
probability  of a point process is equivalent to a conditional probability, and
 $\mathbb{P}^{!o}$ denotes the probability conditioned on there being a point of
 the process at the origin but not including the point (the point at the origin is the receiver, which
 of course does not contribute to the interference). 
Since $\S$ is exponentially distributed, the success probability is given by
\begin{equation}
  \p_\eta = \E\exp\left(-\T \I \right),
  \label{eq:succ_main}
\end{equation}
where for notational convenience we have used $\I$ to denote $\I(o)$. We will use the
standard asymptotic notation $O(\cdot)$, $o(\cdot)$, $\Omega(\cdot)$, $\Theta(\cdot)$, and $\sim$, always taken
as $\eta\to 0$ (unless otherwise noted).

\section{Outage Probability Scaling at Low Interferer Density}
\label{sec:outage_prob}
\subsection{General result}
\label{sec:section_rayleigh}
In this section we show that for a wide range of MAC protocols,
\begin{equation}
  \p_\eta \sim 1- \gamma \eta^{\kappa}, \quad \eta \rightarrow 0.
  \label{eq:claim}
\end{equation}
 While the spatial contention $\gamma$ depends on $\theta$, $\alpha$, and the MAC scheme,
 the interference scaling exponent $\kappa$ 
 depends on $\alpha$ and the MAC, but not on $\theta$. When $\Phi$ is a homogeneous Poisson
 point process (PPP) of intensity $1$, for example, and ALOHA  with
 parameter $\eta\leq1$ is used as the MAC, the success probability
 is \cite{net:Baccelli06}
 \begin{equation}
  \p_\eta=\exp\left(-\eta\int_{\R^2}\frac{1}{ 1+\T^{-1}\l(x)^{-1}}\d x \right)\,.
  \label{ps_ppp}
\end{equation}
 Hence for small $\eta$,
 \[\p_\eta\sim 1-\eta \underbrace{\int_{\R^2}\Delta(x)\d x}_\gamma,\]
 where 
 \begin{equation}
   \Delta(x)= \frac{1}{1+\T^{-1}\l(x)^{-1}}\,.
 \label{delta}
\end{equation}
 Hence  $\kappa=1$ for a PPP with  ALOHA.
The parameter $\kappa$ 
 indicates the gain in link performance when
 the density of transmitters is decreased. More precisely if $\kappa>1$, it is
easy to observe that 
\[\frac{\d \p_\eta}{\d \eta}\Big|_{\eta =0}  = 0.\] 
So for $\kappa>1$, the network can accommodate a certain density of interferers
with negligible effect on the outage, while for $\kappa=1$, when increasing the density
from $0$ to $\d \eta$, the success probability decreases
by  $\gamma\d \eta$.

We begin by  proving that the exponent $\kappa$ cannot take arbitrary values.
Let $\mathcal{K}_\eta(B)$, $B \subset \R^2$, denote the second-order reduced moment measure, defined as
the expected number of points of $\Phi_\eta$ in $B$, given that there is a point at the origin but not counting
that point, normalized by the density of the process:
\[\mathcal{K}_\eta(B)\triangleq \lt^{-1}\E\sum_{\x \in \G} \i(\x \in
B).\]
Alternatively, $\mathcal{K}_\eta$ can be expressed as
\[\mathcal{K}_\eta(B) =  \lt^{-2}\int_B \rho^{(2)}_\eta (x)\d x,\]
where $\rho^{(2)}_\eta(x)$ is the second-order product density of $\Phi_\eta$ \cite{net:Stoyan95,net:Daley-VereJones07}.
For motion-invariant point processes, $\rho^{(2)}(x)$ is a function of $\|x\|$ only, so we may use $\rho^{(2)}(r)$ instead,
for $r\in\R^+$.
Intuitively $\rho^{(2)}(r)\d x \d y$ represents the probability of finding two
points of the process located at $x$ and $y$ with $\|x-y\|=r$.  The second-order
measure $\mathcal{K}_\eta(B)$ is a positive and positive-definite (PPD) measure
\cite{net:Daley-VereJones07}, and hence it follows that
\begin{equation}
 \mathcal{K}_\eta(B+x) < C_B(\eta),\quad \forall x \in \R^2,
 \label{eq:kbound}
 \end{equation}
whenever $\mathcal{K}_\eta(B)<\infty$, where $C_B(\eta)<\infty$ is a constant that does not depend on $x$.
Specializing $\mathcal{K}_\eta(B)$ to the case of a disk centered at the origin, we obtain
Ripley's K-function, defined as $ K_\eta(R) =\mathcal{K}_\eta(B(o,R))$, which, \chr{ when multiplied by $\lt$}, denotes the average number of points in a ball of
radius $R$ conditioned on there being a point at the origin but not counting it. The K-function is often more
convenient to use and sufficient to characterize second-order statistics relevant for motion-invariant point process.
For \chr{a Poisson point process $K_\eta(R)=\pi R^2$, and for} any stationary point process, $K_\eta(R)\sim \pi R^2$ as $ R\rightarrow
\infty$ \cite{net:Stoyan95}.
\begin{theorem}[Bounds on the interference scaling exponent $\kappa$]
  \label{thm:one}
  Any \ch{slotted} 
    MAC protocol that results in a motion-invariant transmitter set
  of density $\eta$ such
  that\footnote{See the discussion after the proof.},
  \begin{equation}
     \lim_{\eta \rightarrow 0} \mathcal{K}_\eta(S_1) < \infty,
	  \quad\text{where } S_1=[0,1]^{2},\tag{C.1}
	  \label{conda}
  \end{equation}
 has the interference scaling exponent  
 \[1\leq \kappa.\]
If for the MAC protocol there exists a $R>0$ such that 
\begin{equation}
 \lim_{\eta \rightarrow 0}\eta K_\eta (R\eta^{-1/2}) >0 \,,\tag{C.2}
 \label{condb}
\end{equation}
%
then 
\[\kappa \leq \alpha/2.\]
  \label{thm:zero}
\end{theorem}
Proof: See Appendix A.

{\em Discussion of the conditions:}
\begin{enumerate}
  \item  \ch{ For a set $B \subset \R^2$, let $\Phi(B)$ denote the number of points of $\Phi\cap B$.} 
           Using a similar
	  argument as in the proof, it is easy to observe that
	  \[\EP[\Phi(B(o,R))] = \lambda\eta \mathcal{K}_\chr{\eta}(B(o,R)) < \lambda\eta \ceil{\pi R^2}\mathcal{K}_\eta (S_1).\] 
		  Condition \eqref{conda} states that \ch{$\lim_{\eta \rightarrow 0} \mathcal{K}_\eta(S_1)<\infty$}, which implies
	  \[\EP[\Phi(B(o,\eta^{-a})] \rightarrow 0\qquad\text{for }a<1/2\,.\]
	  \eqref{conda} implies that the average number of points in a
	ball of radius $R_\eta=\eta^{-a}$, $a<1/2$, goes to
	zero as the density tends to zero. This condition is violated when the
	average nearest-interferer distance remains constant with decreasing density $\eta$.
	For example consider a cluster point
	process with cluster density $\eta$ and a constant mean number of nodes per cluster
	(see Subsection \ref{sec:parent} for a detailed discussion of this example).
	In this case, Condition \eqref{conda} is violated.
 \item Since $\lambda \eta K_\eta(R \eta^{-1/2})$ is the average number of
points in a ball of radius $R \eta^{-1/2}$, Condition \eqref{condb}
	requires the number of points inside a ball of radius
	$R\eta^{-1/2}$ to be greater than zero. 
	By the sphere-packing argument, in any stationary
point  process of density $\lambda$, the probability that the nearest neighbor 
 is within a distance $\sqrt{2/\sqrt{3}}\lambda^{-1/2}$ is greater than
zero. In other words, the probability of the event that all the nearest neighbors are further away 
than $1.075\eta^{-1/2}$ is zero. Hence $\eta\lambda
K_\eta(R\eta^{-1/2} ) >0$, where $R=\sqrt{\frac{2}{\sqrt{3}}}$.  But this does
not strictly satisfy Condition \eqref{condb} which requires the {\em limit} to be greater than zero.
 Except possibly for pathological
	cases\footnote{We are not aware that any such case exists.}, this condition is generally valid since the nearest-neighbor
distance scales (at most) like $\eta^{-1/2}$ when the point process has
	density $\eta\lambda$.
	So,   while Condition \eqref{conda} requires the nearest-interferer distance to increase with decreasing
	$\eta$, Condition \eqref{condb} requires an interferer to be present
	at a distance $\Theta(\eta^{-1/2})$. 
Any MAC that schedules
	the nearest interferer at an average distance that scales with $\eta^{-1/2}$
	satisfies these two conditions, and in this case, $1\leq \kappa
	\leq \alpha/2$.
  \item  Indeed if Condition \eqref{conda} is violated then
\[ \ch{\p_0\triangleq} \lim_{\eta\rightarrow 0}\p_\eta<1.\]
\end{enumerate}
Based on this discussion, we can define the class of {\em reasonable} MAC schemes:
\begin{definition}
A {\em reasonable} MAC scheme is a MAC scheme for which Conditions \eqref{conda} and \eqref{condb} hold.
\end{definition}
Theorem \ref{thm:one} implies that for all reasonable MAC schemes, $1\leq \kappa \leq \alpha/2$.
A MAC scheme for which $\lim_{\eta\rightarrow 0} \p_\eta<1$ would clearly be unreasonable---it
would defeat the purpose of achieving high reliability as the density of interferers is decreased.

\subsection{Achieving the boundary points $\kappa=1$ and $\kappa=\alpha/2$}
In this section, we provide 
exact conditions on the MAC protocols which achieve the boundary points
$\kappa=1$ and $\kappa=\alpha/2$.
ALOHA is a simple MAC protocol, and its fully distributed nature makes it very
appealing. As shown before Poisson
distribution of transmitters with ALOHA as the MAC protocol has $\kappa=1$. 
ALOHA with parameter $\eta$ leads to independent thinning
of $\Phi$, and the resultant process has an average nearest-neighbor distance that scales like $1/\sqrt{\eta\lambda}$.
Independent thinning of  a point process  does  not guarantee that there is no
receiver within a distance  $R=c/\sqrt{\eta\lambda}$, $c<1$, as $\eta\to 0$.
If suppose there are $n$ points originally in the ball $B(o,R)$, the probability
that none of the points are selected by ALOHA is $(1-\eta\lambda)^{n}$. So although
ALOHA with parameter $\eta$ guarantees an average nearest-neighbor distance
$(\eta\lambda)^{-1/2}$, there is a finite probability $1-(1-\eta\lambda)^{n}$ that the ball
$B(o,R)$ is not empty. So essentially ALOHA leads to a {\em soft} minimum
distance proportional to $(\eta\lambda)^{-1/2}$,  and as we state in the following theorem,  results in
$\kappa=1$ for any network with ALOHA.
\begin{theorem}[Achieving $\kappa=1$]
  \label{thm:aloha}
Let the MAC scheme be ALOHA with transmit probability $\eta$. If 
\begin{equation}
  \int_{\R^2}\int_{\R^2}
  \rho^{(3)}(x,y)\Delta(x)\Delta(y)\d x\d y <\infty, \tag{C.3}
  \label{eq:main_cons}
\end{equation}
the outage probability satisfies
\begin{equation} 
  \label{eq:intscal}
\p_\eta\sim 1-\gamma \eta, \quad \eta\rightarrow 0,
\end{equation}
where 
\begin{equation}
\gamma=\lambda^{-1}\int_{\R^2} \rho^{(2)}(x)\Delta(x)\d x. 
\label{gamma_thm2}
\end{equation}
$\rho^{(3)}(x,y)$ denotes the third-order product
density  \cite{net:Hanisch83,net:Stoyan95,net:Daley-VereJones07}
of the point process $\Phi$.
\label{stat:ps1}
\label{stat:IEEEproof_ps1}
\end{theorem}
\begin{IEEEproof}
 For ALOHA, the resulting transmitter process $\G$ is
 an independently thinned version of $\Phi$. From \eqref{eq:succ_main} we obtain
  \begin{eqnarray} 
	\p_\eta&=&\E\exp\left(-\theta\sum_{\x\in\G}\h{\x}\l(\x)\right)\IEEEnonumber\\
	&=&\E\exp\left(-\theta
	\sum_{\x\in\Phi}\h{\x}\l(\x)\mathbf{1}(\x\in\G)\right)\IEEEnonumber.
  \end{eqnarray}
  Since $\h{\x}$ is exponential,
  \begin{eqnarray}
	\p_\eta&=&\E\prod_{\x\in\Phi}\frac{1}{1+\theta\mathbf{1}(\x\in\G)\l(\x)}.
	\label{eq:main_err}
  \end{eqnarray}
  It is easy to observe that 
  \[ \frac{1}{1+\theta\mathbf{1}(\x\in\G)\l(\x)}=
  \frac{\mathbf{1}(\x\in\G)}{1+\theta\l(\x)}+1-\mathbf{1}(\x\in\G).\]
  Averaging over the ALOHA MAC  yields
  \[\chr{\mathbb{E}}\left[
  \frac{1}{1+\theta\mathbf{1}(\x\in\G)\l(\x)}\right] =
  1-\eta\Delta(\x)\]
  Hence we obtain 
  \begin{align*}
   \p_\eta &= \E\left[\prod_{\x \in \Phi} 1-\eta\Delta(\x)
  \right]\\
    &= \mathcal{G}^{!o}\left[1-\eta \Delta(\cdot) \right]\,,
  \end{align*}
  where   $\mathcal{G}^{!o}$ is the reduced probability generating functional. 
  Proceeding as in \cite[Thm. 9.6.5]{net:Daley-VereJones07}, it follows
  that 
  \[ \p_\eta =1-\eta \lambda^{-1}\int_{\R^2}
  \rho^{(2)}(x)\Delta(x)\d x +o(\eta), \]
  when \eqref{eq:main_cons} is satisfied.
\end{IEEEproof}
Condition \eqref{eq:main_cons} essentially bounds the second
moment of $\sum_{\x \in \Phi}\Delta(\x)$. It plays a similar role as  the third
moment constraint in the Berry-Ess\'een theorem. 

\begin{figure}
\begin{center}
\subfigure[$\eta=1/9$]{\label{fig:TDMA9}\includegraphics[width=0.8\columnwidth,height=.75\columnwidth]{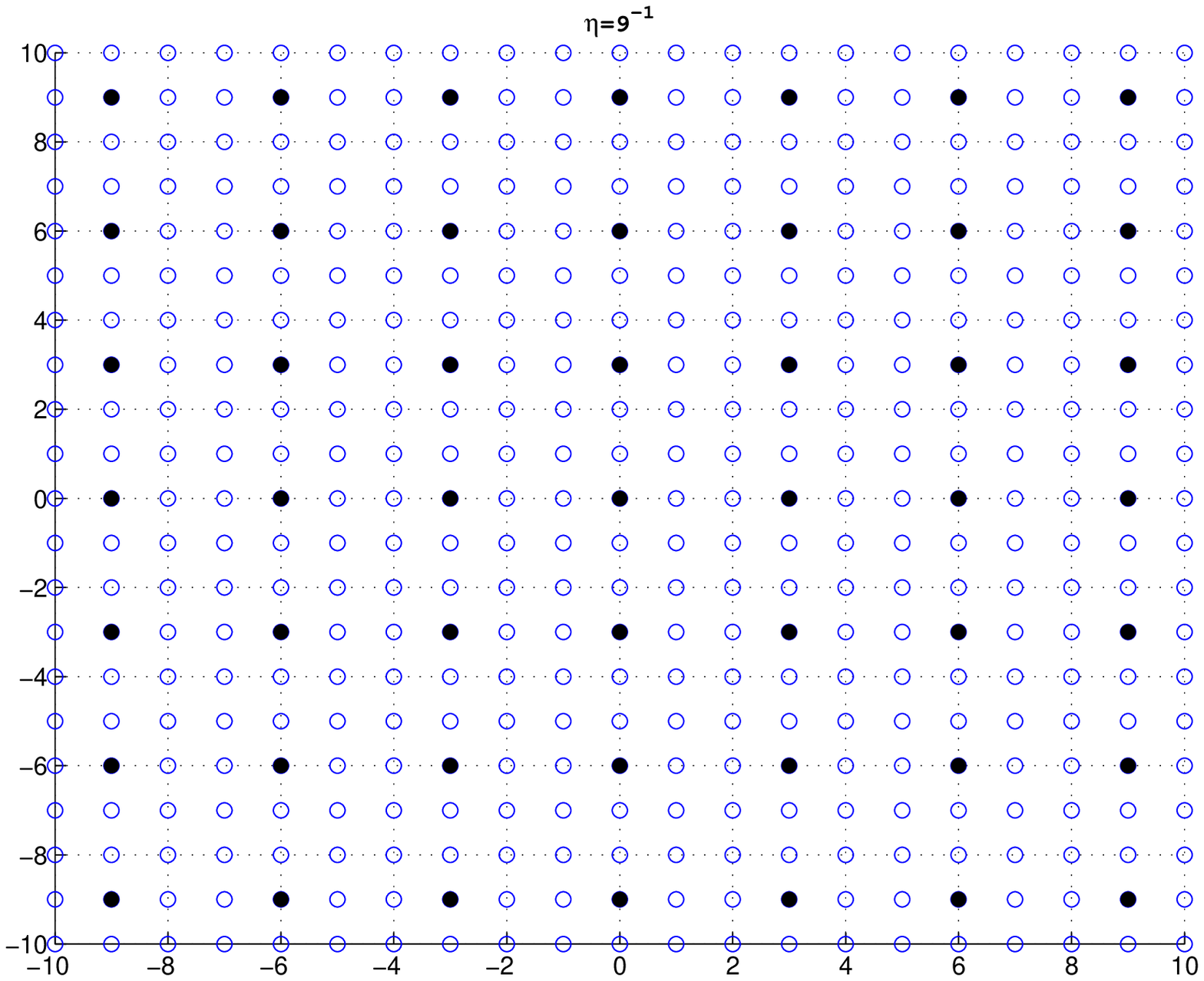}}                
\subfigure[$\eta=1/16$]{\label{fig:TDMA16}\includegraphics[width=0.8\columnwidth,height=.75\columnwidth]{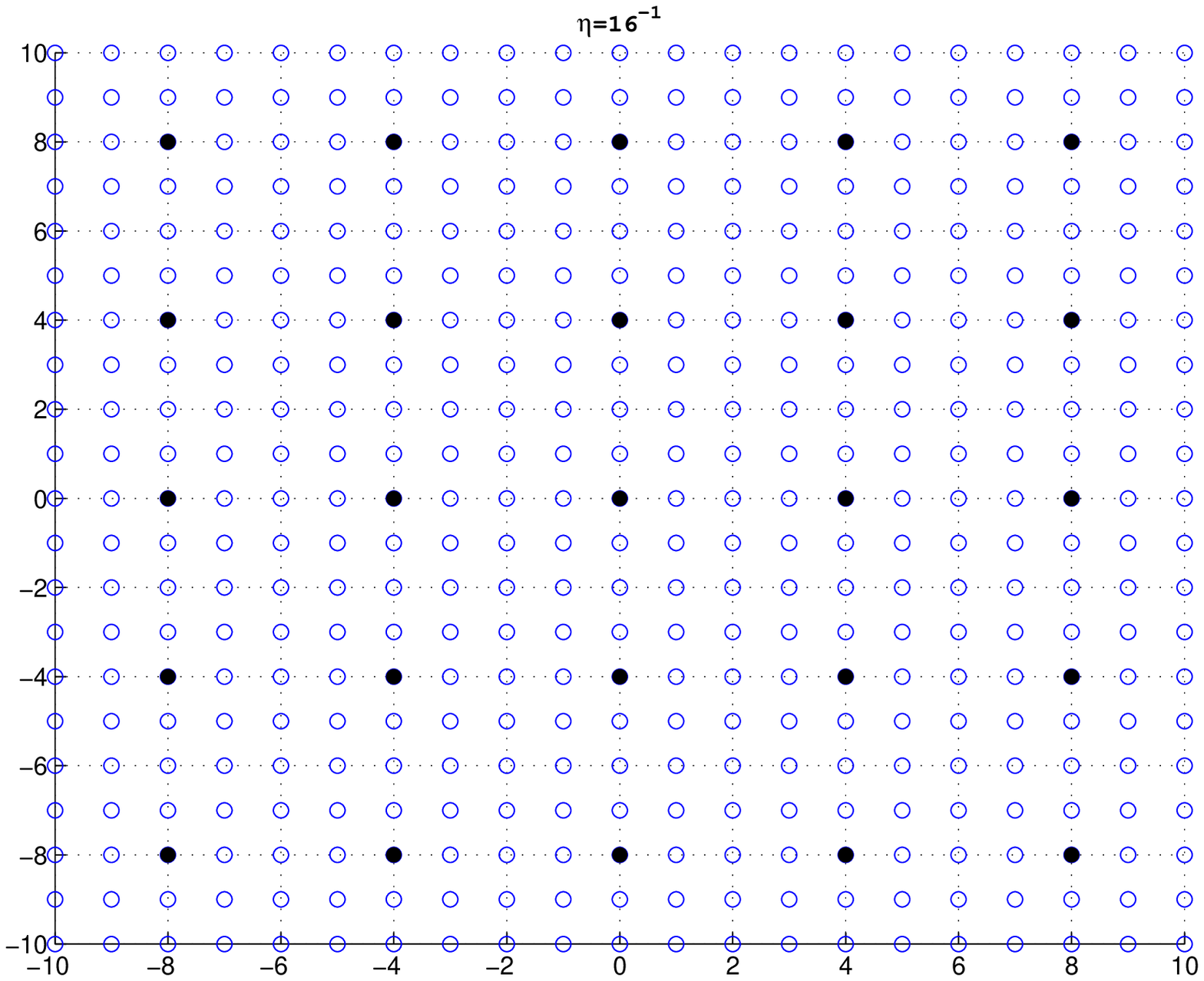}}
\end{center}
\caption{{\em Reasonable} TDMA on a two-dimensional lattice $\Z^2$ for $\eta=1/9$ and $\eta=1/16$. In this
arrangement, the nearest interferer is at distance $\eta^{-1/2}$.}
\label{fig:TDMA1}
\end{figure}

\begin{figure}
\centering
\subfigure[$\eta=1/9$]{\label{fig:aTDMA9}\includegraphics[width=0.8\columnwidth,height=.75\columnwidth]{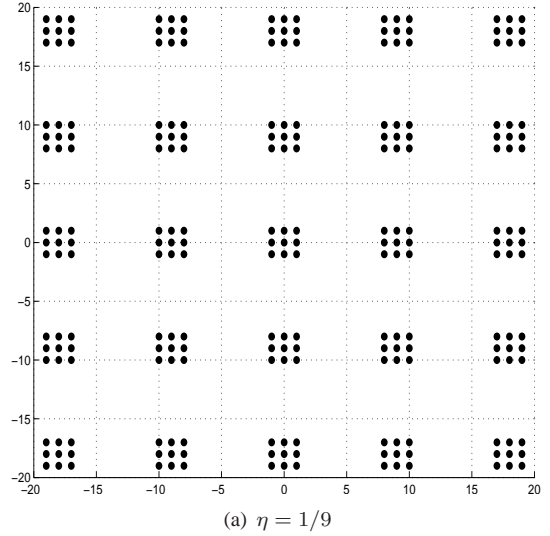}}                
\subfigure[$\eta=1/16$]{\label{fig:aTDMA16}\includegraphics[width=0.8\columnwidth,height=.75\columnwidth]{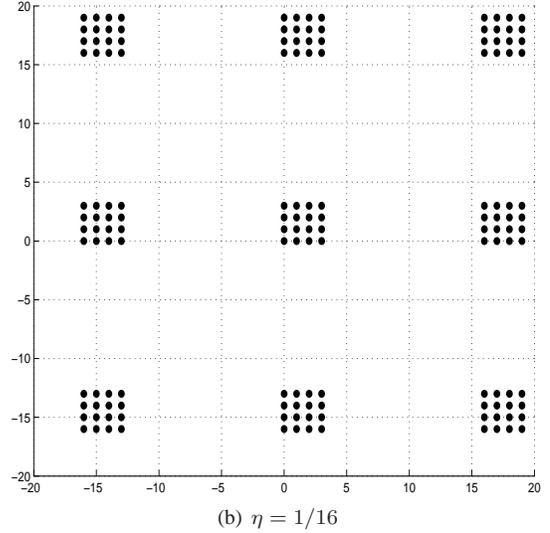}}
\caption{{\em Unreasonable} TDMA on a two-dimensional lattice $\Z^2$ for $\eta=1/9$ and $\eta=1/16$. In this
case, the nearest interferer is at unit distance --- irrespective of $\eta$.}
\label{fig:TDMA2}
\end{figure}

In the previous theorem, we characterized the scaling law for ALOHA,
where only the {\em average} distance to the nearest interferer scales
as $\eta^{-1/2}$. We now consider the MAC protocols in which the nearest
interferer distance scales as $\eta^{-1/2}$ {\em almost surely}.
For example, a TDMA scheme in
which the distance between the nearest transmitters scale like $\eta^{-1/2}$
falls into this category. Figures \ref{fig:TDMA1} and \ref{fig:TDMA2} illustrate two 
different TDMA scheduling schemes on a lattice network.
In the scheme in Figure \ref{fig:TDMA1}, we observe that
$\rho^{(2)}_\eta(x/\sqrt{\eta})=0$ for $x <1$, while this is not the case
in the modified TDMA scheme in Figure \ref{fig:TDMA2}. More precisely, it is easy to
observe that the minimum-distance criterion
\begin{equation}
  \lim_{\eta \rightarrow 0}\int_0^\infty \eta^{-2}\rho^{(2)}_\eta(r\eta^{-1/2})
  r^{1-\alpha}\d r <\infty.
  \label{eq:cond1}
\end{equation}
holds in the TDMA scheme illustrated in Figure \ref{fig:TDMA1} but not in the alternative {\em unreasonable}
TDMA of Figure \ref{fig:TDMA2}. The factor $\eta^{-2}$ in front of the
 second-order product density is required since $\rho_\eta^{(2)}(x)$ scales as $\eta^{2}$. In
 the first TDMA scheme, we also observe that the resulting transmitter process is
 self-similar if both axes are scaled by $\eta^{-1/2}$. In contrast, in the unreasonable TDMA version, the
 nearest-interferer distance stays constant with decreasing $\eta$. 
 
   Next we state the conditions for $\kappa=\alpha/2$.
 

\begin{theorem}[Achieving $\kappa=\alpha/2$]
  \label{thm:two}
  Consider a motion-invariant point process $\Phi$ and a MAC protocol
  for which the following three conditions are satisfied:
    \begin{align}
	\gamma\triangleq  &\lim_{\eta \rightarrow 0} \T \lambda^{-1}\eta^{-\alpha/2-1} \int_{\R^2}
	  \l(x)\rho_\eta^{(2)}(x)\d x \in (0,\infty) \tag{C.4}
    \label{cond31}\\
   & \int_{\R^2}  \l^2(x)\rho_\eta^{(2)}(x)\d x = o(\eta^{\alpha/2+1}) \tag{C.5}
     \label{cond32}\\
  & \int_{\R^2} \l(x)\l(y)\rho_\eta^{(3)}(x,y)\d x \d y = o(\eta^{\alpha/2+1})  \, \tag{C.6}
     \label{cond33}
  \end{align}   
Then 
\[\lim_{\eta\rightarrow 0}\frac{1-\p_\eta}{\eta^{\alpha/2}} =\gamma\,. \]
\end{theorem}
Proof: See Appendix B.

The conditions provided in Theorem \ref{thm:two} will be satisfied by most MAC protocols in which the
nearest-interferer distance scales like $\eta^{-1/2}$ a.s.  Using the substitution $x\rightarrow \eta^{-1/2}y$,
Condition \eqref{cond31}
	can be rewritten as
	\[ \int_{\R^2}
	\eta^{-\alpha/2}\l(\eta^{-1/2}y)\eta^{-2}\rho^{(2)}_\eta(y\eta^{-1/2})\d y.\]
	Due to the factor $\eta^{-\alpha/2}$ in front of $\l(y\eta^{-1/2} )$,
	we can immediately see that 
	\[\eta^{-\alpha/2}\l(y\eta^{-1/2} ) \rightarrow \|y\|^{-\alpha}, \]
  and hence only the tail behavior of the path loss model matters. We
  also observe that for $\gamma$ in Condition \eqref{cond31} to be finite,
  $\eta^{-2}\rho^{(2)}_\eta(\eta^{-1/2}y)$ should decay
   to zero in the neighborhood of $\|y\|=0$. This observation leads
  to the following corollary about the CSMA protocol.

\begin{corollary}[CSMA]
  \label{cor:CSMA}
  Any MAC protocol which selects a \ch{motion-invariant} transmitter set $\G$ of
	density $\eta \lambda$ such
	that the second-order product density $\rho_\eta^{(2)}(x)$ is zero  for
	$x\in B(o,c\eta^{-1/2})$ for some $c>0$,
	and satisfies Condition \eqref{cond33} has the interference scaling exponent $\kappa
	=\alpha/2$.
\end{corollary}
Proof: See Appendix C.

The following corollary states that the bounds easily extend to $d$-dimensional networks.
The proof techniques are the same.
\begin{corollary}[$d$-dimensional networks]
  \label{cor:ddim}
  Consider a $d$-dimensional motion-invariant point process $\Phi$ of intensity $\lambda$, and a MAC scheme
  that, as a function of a thinning parameter $\eta$, produces motion-invariant point processes $\Phi_\eta$ of
  intensity $\lt=\eta\lambda$.
  If Condition \eqref{conda} in Theorem \ref{thm:one} holds with $\R^2$ replaced by $\R^d$, and Condition
  \eqref{condb} holds with $\eta^{-1/2}$ replaced by $\eta^{-1/d}$, then
  \[ \p_\eta\sim 1-\gamma\eta^\kappa\,, \qquad\text{as }\;\eta\rightarrow 0\,, \]
  where $1\leq \kappa \leq \alpha/d$.
\end{corollary}
It can be seen that the condition $\alpha>d$, required for finite interference a.s.$\,$\ch{\cite{net:Haenggi08now}}, is reflected in
these bounds. If $\alpha<d$, the set of possible $\kappa$ is empty.

In the next section we consider networks with different spatial node distributions
and MAC protocols and verify the theoretical results by simulations.
\section{Examples and Simulation Results}
\label{sec:examples}
\subsection{Poisson point process (PPP) with ALOHA}
\label{sec:PPP}
When $\l(x)=\|x\|^{-\alpha}$, the success probability in a PPP is well studied
\cite{net:Baccelli06,net:Haenggi08now,net:Weber05tit}; it has been shown to be
\[\p= \exp\left(
-\lambda\T^{2/\alpha}\frac{2\pi}{\alpha}\Gamma(2/\alpha)\Gamma(1-2/\alpha)
\right).\]
When ALOHA  with parameter $\eta$ is used as the MAC protocol, the resulting
process is also a  PPP with density $\eta \lambda$ and hence the success
probability is
\[\p_\eta= \exp\Bigl(
-\eta \underbrace{\lambda\T^{2/\alpha}\frac{2\pi}{\alpha}\Gamma(2/\alpha)\Gamma(1-2/\alpha)}_\gamma
\Bigr). \]
From the above expression we observe that $\p_\eta \sim 1-\eta\gamma$ as
$\eta\rightarrow 0$. For a
PPP, $\rho^{(2)}(x)=\lambda^2$, and it can be verified that
\[ \gamma=\lambda^{-1} \int_{\R^2} \frac{\rho^{(2)}(x)}{1+\theta^{-1}\|x\|^{\alpha}}\d x\,,
 \]
 in accordance with \eqref{gamma_thm2} in Theorem \ref{thm:aloha}.

\subsection{Hard-core processes}
\label{sec:Outage for Hard-core processes}
Hard-core point process possess a minimum distance between the points and hence
are useful in modeling CSMA-type MAC protocols. Hard-core
processes exhibit an intermediate regularity level between the Poisson point
processes and lattice processes. A good model for CSMA are
Matern hard-core processes of minimum distance $h$, obtained
 by dependent thinning of a PPP  as follows \cite[pp.~162]{net:Stoyan95}:
 Each node of the PPP is marked independently with a uniform random number
 between $0$ and $1$. A node $x$  with mark $m(x)$ is retained if the ball $B(x,h)$
 contains no other nodes of with a mark less than $m(x)$. Starting with a
 PPP of intensity $\lambda_p$, this leads to a stationary point process of density 
\begin{equation} \label{eq:lambdahd}
\lambda'=\frac{1-\exp(-\lambda_{p} \pi h^2)}{\pi h^2}\,.
 \end{equation}
  \chr{
 Let  $\VV(x_1,\hdots,x_n)$ denote the area of the intersection of discs of radius $h$ centered around $x_i$, with the convention $\VV(x_1)=\pi h^2$.  Also define
 \begin{multline*}
  f(m_1,\hdots,m_k)=\\
   \exp\left(-\lambda_p\sum_{J\subset\{1,\hdots,k\}}(-1)^{|J|+1}m_{\min\{J\}}\VV(J)\right)\,,
 \end{multline*}
 where $\VV(J)=\VV(x_{a_1},\hdots,x_{a_{|J|}})$, when $ J=\{a_1,\hdots,a_{|J|}\}$.
 Then the $n$-th order product density of the Matern hard-core process \cite{net:Baccelli09now2} is
\begin{multline}
\rho^{(n)}(x_1,\hdots,x_{n-1})=\\
\begin{cases}
0,  \qquad\text{if } \|x_i-x_j\|<h,  \text{for any } i,j, &\\
  n!\lambda_p^n\int_A f(m_1,\hdots,m_n)\d m_1...\d m_n, & \text{otherwise}\,,
 \end{cases}
 \label{eq:rho_mat}
 \end{multline}
where $A$ is the subset of $[0,1]^n$ where $0\leq m_1\leq \hdots \leq m_n \leq 1$.}
The second-order product density can be easily obtained from \eqref{eq:rho_mat} to be $\rho^{(2)}(x)=$
\begin{equation}
\begin{cases}
  0 &  r<h\\
  \frac{2\Gamma_h(\|x\|)(1-\exp(-\lambda_p c))-2c(1-\exp(-\lambda_p\Gamma_h(\|x\|)))}{c\Gamma_h(\|x\|)(\Gamma_h(\|x\|)-c)} &  h\leq \|x\|\leq2h\\
  \lambda'^2 & \|x\|>2h
\end{cases}
\label{eq:rho2hardcore1}
\end{equation}
where $c=\pi h^2$ and
\[ \Gamma_h(r)=2\pi h^2- 2h^2\arccos\left(\frac{r}{2h}\right)+\frac{r}{2}\sqrt{4h^2-r^2}\,. \]
 This second-order product density can also be found in \cite[p.~164]{net:Stoyan95}. \ch{Since the point process
 is motion-invariant, $\rho^{(2)}(x)$ only depends on the magnitude of $x$.}


\subsubsection{Hard-core process with ALOHA}
\begin{figure}
\centering
\includegraphics[width=.9\columnwidth]{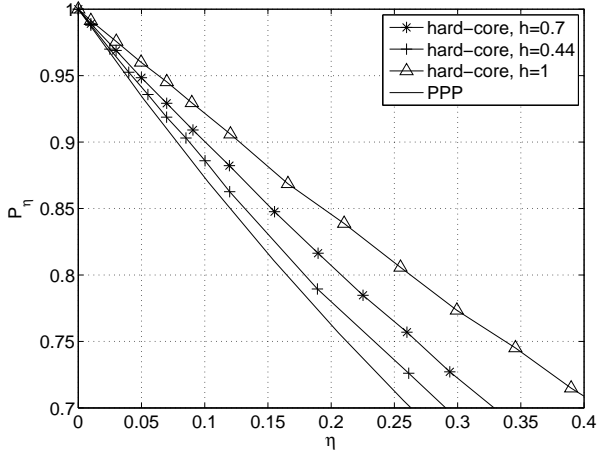}
\caption{Simulation of the outage probability of a hard-core point process with ALOHA for a path loss law
$\l(x)=\|x\|^{-4}$ and $\theta=2$. The hard-core
radii are $h=0$ (PPP), $h=0.44$, $h=0.7$, and $h=1.0$.
The density of the underlying PPP $\lambda_p$ is adjusted according to \eqref{eq:lambdahd}
so that in all cases, $\lambda=0.194$.
It is observed that the decay for small $\eta$ is indeed linear, as predicted by $\kappa=1$.}
\label{fig:ALOHA}
\end{figure}

In this case, we start by generating a hard-core point process with intensity $\lambda=\lambda'$ as given in
\eqref{eq:lambdahd} and then apply independent thinning with probability $\eta$.
In Figure \ref{fig:ALOHA} the success probability in a hard-core process network
with ALOHA is shown. As proved in Theorem \ref{thm:aloha}, we
observe that $\p_\eta\sim 1-\gamma\eta$, where $\gamma$ is given by
\eqref{gamma_thm2}.
It can be observed that  the outage probability improves as $h$ increases, as expected.

\begin{figure}
\centering
\includegraphics[width=.9\columnwidth]{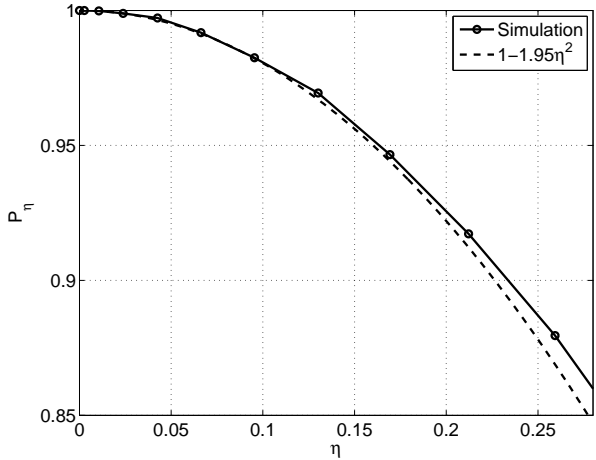}
\caption{Simulation of the success probability of a PPP of intensity $\lambda=0.3$
with CSMA, where the hard-core distance $h$ is adjusted to
achieve intensity $0.3\eta$, for a path loss law
$\l(x)=\|x\|^{-4}$ and $\theta=2$. For comparison, the curve $1-\gamma\eta^2$ is also shown. $\gamma\approx 1.95$
is obtained from Cor.~\ref{cor:matern_csma}. It is observed
that the decay for small $\eta$ is indeed quadratic, as predicted by $\kappa=\alpha/2=2$.}
\label{fig:ppp_csma}
\end{figure}

\subsubsection{Poisson point process with CSMA}
In this case, we start from a PPP with intensity $\lambda$ and adjust $h$ to thin the
process to a Matern hard-core process of intensity $\lt=\lambda\eta$, as given by \eqref{eq:lambdahd}
(where now $\lambda_p:=\lambda$ and the resulting $\lambda'$ is the transmitter density $\lt$). 
Since we are interested in small $\eta$ or, equivalently,
large $h$, we obtain from \eqref{eq:lambdahd} that
$h\to \sqrt{1/(\pi\eta\lambda)}$  for large $h$.
So we have that the second-order product
density $\rho^{(2)}(x)$ is zero for $\|x\| < \sqrt{1/(\pi\eta\lambda)}$, \chr{ and Condition (C.6) can be verified using \eqref{eq:rho_mat},} and hence
the conditions of Corollary \ref{cor:CSMA} are satisfied. Hence scaling the inhibition radius with $\eta^{-1/2}$ between the transmitters
leads to $\kappa =\alpha/2$, and the constant $\gamma$ is given by the
following corollary.
\begin{corollary}
  \label{cor:matern_csma}
 When the transmitters are modeled as a Matern hard-core process and the MAC
 protocol decreases the density by increasing the inhibition radius $h$ such that
 $h=\Theta(\eta^{-1/2})$, the spatial
 contention parameter $\gamma$  
 is given by
 \begin{equation}
\gamma=\frac{\theta\lambda^{\alpha/2}\pi^{\alpha/2}2^{3-\alpha}}{\alpha-2}+4\theta\lambda\pi^{2}\int_{1/\sqrt{\lambda\pi}}^{2/\sqrt{\lambda\pi}}\frac{r^{1-\alpha}}{g(r)}\d r\,,
\label{gamma_hc}
\end{equation}
where\[
g(r)=2\pi-2\arccos\left(\frac{\sqrt{\lambda\pi}}{2}r\right)+\frac{r\sqrt{\lambda\pi}}{2}\sqrt{4-\pi\lambda r^{2}}.\]
\end{corollary}
\figref{fig:ppp_csma} shows a simulation result for a PPP of intensity $\lambda=0.3$, where
hard-core thinning with varying radius $h$ is applied to model a CSMA-type MAC scheme, for $\alpha=4$.
It can be seen that the outage increases indeed only quadratically with $\eta$ and that the
asymptotic expression provides a good approximation for practical ranges of $\eta$. The spatial contention
$\gamma$ can be obtained from \eqref{gamma_hc}; it is $\gamma\approx 1.95$.

  \subsection{Poisson cluster processes \chr{(PCP)}}
\label{sec:Outage for Cluster processes}
A Poisson cluster process \cite{net:Stoyan95, net:Daley-VereJones07}  consists of
the union of finite and independent daughter point processes (clusters) centered at parent points that form a PPP.
The parent points themselves are not included in the process.
 Starting with a parent point process of  density $\mu$ and deploying 
$c$ daughter points per parent on average, the resulting cluster process has a density of  $\lambda=\mu c$.
The success probability in a Poisson cluster process, when the
number of daughter points in each cluster is a Poisson random variable with mean
$c$ is  \cite{net:Ganti09tit}
\begin{multline}
\p=
\exp\left(-\mu\int_{\R^{2}}\Big[1-\exp(-c\beta(y))\Big]\d y\right)\\
 \times\int_{\R^{2}}\exp(-c\beta(y))f(y)\d y,\label{eq:cluster_prob}
 \end{multline}
 where
 \begin{equation*}
   \beta(y)=\int_{\R^{2}}\Delta(x-y)f(x)\d x,
  \end{equation*}
   and $f(x)$ is the density function of the cluster with $\int_{\R^2}f(x) \d x=1$.
    In a  Thomas cluster process each point is scattered using a symmetric
normal distribution with variance $\sigma^{2}$ around the parent.
So the density function $f(x)$ is given by\[
f(x)=\frac{1}{2\pi\sigma^{2}}\exp\left(-\frac{\Vert x\Vert^{2}}{2\sigma^{2}}\right).\]
The second-order product density for a Poisson cluster process is
\cite{net:Stoyan95,net:Daley-VereJones07,net:Haenggi08now}
\[ \rho^{(2)}(z)=\lambda^2\left[1+\frac{(f*f)(z)}{\mu}\right],\]
which, for 
a Thomas cluster process, evaluates to  
\begin{equation}
\rho^{(2)}(x)=\lambda^2\left(1+\frac{1}{4\pi\sigma^2\mu }\exp\left(-\frac{\|x\|^2}{4\sigma^2}\right)\right).\label{eq:clustrho2}
\end{equation}

\subsubsection{ALOHA (daughter thinning)}
\begin{figure}
\centering
\includegraphics[width=.9\columnwidth]{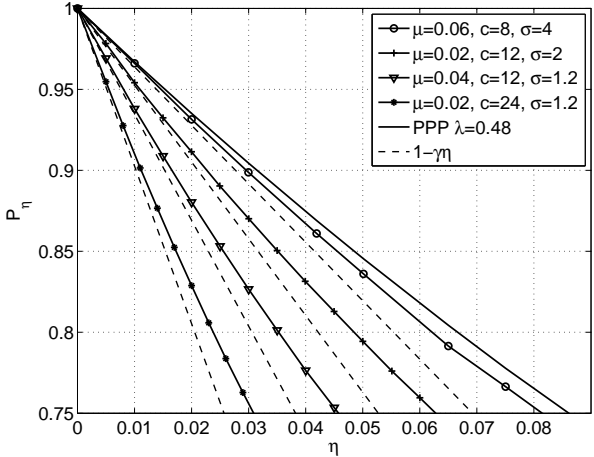}
\caption{Simulation of ALOHA on Poisson cluster networks with four different parameters $\mu$ (parent density) and $c$
(mean number of point per cluster).
In all cases, $\alpha=4$, $\theta=2$, and the intensity of the cluster process is $0.48$. The curve for the PPP is also
shown for comparison.
The dashed lines are the asymptotic expressions $\p_\eta\approx 1-\gamma\eta$ for the four clustered cases.
The values of $\gamma$ are $3.61$, $4.74$, $6.54$, and $9.73$, respectively.}
\label{fig:cluster_ALOHA}
\end{figure}

For ALOHA, each node is retained with a probability
$\eta$ and the resulting process is again a cluster process with daughter
density $c \eta$. Substituting $c \eta$ for $c$ in \eqref{eq:cluster_prob}, it
is easy to verify that  $\kappa =1$ and $\gamma = \lambda^{-1}\int_{\R^2}
\Delta(x)\rho^{(2)}(x)\d x$,  verifying  Theorem \ref{thm:aloha}.
 In \figref{fig:cluster_ALOHA}, various configurations are shown with the
corresponding analytical approximation obtained by the numerical computation of
$\gamma$ using (\ref{eq:clustrho2}), and we observe a close match for small $\eta$.
\subsubsection{Highly clustered MAC (parent thinning)}
\label{sec:parent}
A high clustered MAC can be obtained by thinning the parents (\ie, keeping or removing
entire clusters) instead of the
daughter points. 
This means that all the points induced by a parent point transmit with probability
$\eta$, and all of them stay quiet with probability $1-\eta$.
Such a MAC scheme causes highly clustered transmissions and high spatial contention.
Even when the density of transmitting nodes gets very small, there are always nodes
near the receiver that are transmitting. Such a MAC scheme violates
Condition  \eqref{conda}  in Theorem \ref{thm:zero}:
The condition in this case is equivalent to
\[\mathcal{K}_\eta([0,1]^2) = \int_{[0,1]^2} \frac{\rho^{(2)}(x)}{\lambda^2}\d
x= \int_{[0,1]^2}
\left[1+ \frac{(f*f)(z)}{\eta \mu} \right] \d z,\]
and it follows that $\lim_{\eta \rightarrow 0}
\mathcal{K}_\eta([0,1]^2) \rightarrow
\infty $.
From
\eqref{eq:cluster_prob}, it is easy to observe that
\[\p_\eta\leq \int_{\R^2} \exp(-c \beta(y))f(y)\d y<1,\]
\ie, the probability of success  never reaches one because of the interference
within the cluster. This is an example of an {\em unreasonable} MAC protocol.

\subsubsection{Clustered MAC (parent and daughter thinning)}
\label{sec:cluster2}
\begin{figure}
\begin{center}
\includegraphics[width=.9\columnwidth]{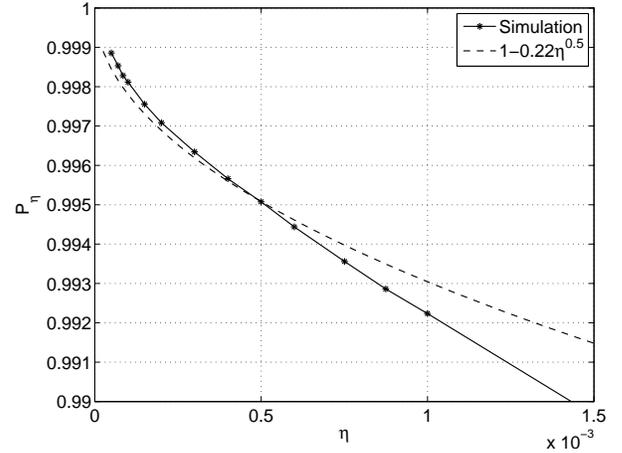}
\centering
\caption{Simulation and analytical asymptotic behavior for a Poisson cluster network with a cluster MAC with
parameter $b=0.5$ for $\alpha=4$, $\theta=2$. The cluster process is a Thomas process with $\mu=0.1$, $c=4$, and
$\sigma=3.6$. It is confirmed that for (very) small $\eta$, the success probability
decreases superlinearly with exponent $\kappa=0.5$.}
\label{fig:clustered-b05}
\end{center}
\end{figure}

The ALOHA and highly clustered MAC schemes can be generalized as follows:
For a MAC parameter $b\in[0,1]$, first schedule entire clusters with probability $\eta^{1-b}$ (parent thinning) and then, within each cluster, each daughter may transmit with probability $\eta^b$ (daughter thinning). This results in a transmitter process $\Phi_\eta$ of intensity
$\eta\lambda$, as desired. It includes ALOHA as a special case for $b=1$, and pure parent thinning for $b=0$. For $b<1$,
the mean nearest-interferer distance scales more slowly than $\eta^{-1/2}$, and we expect $\kappa<1$. Indeed, it follows from
\eqref{eq:cluster_prob} that $\kappa=b$ in this case.
\figref{fig:clustered-b05} shows a simulation result for $b=\kappa=1/2$, which confirms the theoretically predicted sharp
decay of the success probability near $\eta=0$. 

 \subsection{$d$-dimensional TDMA networks}
 In a $d$-dimensional lattice network\footnote{with random translation and rotation for motion-invariance}
  $\Phi=\mathbb{Z}^d$, consider a TDMA MAC scheme, where
 every node transmits once every $m^d$ time slots so that only one node in a $d$-dimensional
 hypercube of side length $m$ transmits. Since it takes $m^d$ time slots for this scheme to give each node one
 transmit opportunity, it is an $m^d$-phase TDMA scheme, and the minimum distance between
 two transmitters is $m$.
In a  regular single-sided one-dimensional $m$-phase TDMA network with Rayleigh
fading  the success probability is bounded as \cite[Eq. (31)]{net:Haenggi09twc}
\begin{equation} \label{eq:bound1}
e^{-\zeta(\alpha)\theta/m^\alpha}\leq \chr{ \p_{1/m}}\leq\frac{1}{1+\zeta(\alpha)\theta/m^{\alpha} },
\end{equation}
where $\zeta$ is the Riemann Zeta function. The following theorem generalizes the bounds to
lattices of dimension $d$. 
 \begin{theorem}
For $m^d$-phase TDMA on $d$-dimensional square lattice networks, the success probability
is tightly bounded as
\begin{equation}
e^{-Z^{(d)}(\alpha)\theta\eta^{\alpha/d}}\leq \p_\eta\leq\frac{1}{1+Z^{(d)}(\alpha)\theta \eta^{\alpha/d}}\,,
 \label{eq:boundd}
 \end{equation} 
 where $\eta=m^{-d}$ and $Z^{(d)}$ is the Epstein Zeta function of order $d$ \cite{net:Epstein03}, defined 
 (in its simplest form) as
 \[ Z^{(d)}(\alpha)\triangleq \sum_{x\in\mathbb{Z}^d\setminus\{o\}} \|x\|^{-\alpha} \,. \]
\end{theorem}
\begin{IEEEproof}
Following the proof of \cite[Prop.~3]{net:Haenggi09twc}, for an $m^d$-phase TDMA network 
\begin{equation} \label{eq:19b}
\p_\eta=\prod_{x\in\mathbb{Z}^d\setminus\{o\}}\frac{(m\|x\|)^\alpha/\T}{1+(m\|x\|)^\alpha/\T}\,,
\end{equation}
Letting $\theta' =\theta/m^\alpha$, we obtain
\begin{equation} 
 \p_\eta^{-1}=\displaystyle\prod_{x\in\mathbb{Z}^d\setminus\{o\}}1+\theta'/\|x\|^\alpha\,.
 \label{ps_inverse}
 \end{equation} 
For
 the upper bound, 
ordering the terms according to the powers of $\theta'$ yields
\[\p_\eta^{-1}=1+\theta'Z^{(d)}(\alpha)+\theta'^2(Z^{(d)}(\alpha)-1)+\ldots\]
Truncating this expansion at the second term, we obtain
  \begin{equation} \label{eq:boundda}
 \p_\eta \leq\frac{1}{1+Z^{(d)}(\alpha)\theta/ m^{\alpha}}\,.
 \end{equation}  
 The lower bound in \eqref{eq:boundd} is obtained by noting that $\eta=m^{-d}$,
 taking the logarithm of \eqref{ps_inverse} and using $\log(1+x)\leq x$. 
   \end{IEEEproof}
Fom the above theorem we observe that for a $d$-dimensional TDMA network,
  \begin{eqnarray} \label{eq:gammadgen}
\gamma^{(d)}&=&Z^{(d)}(\alpha)\theta\\
\kappa^{(d)}&=&\alpha/d,
 \end{eqnarray}
in agreement with
Theorem \ref{thm:two} and Cor.~\ref{cor:ddim}. The conditions of Theorem \ref{thm:two} are satisfied
since the support of $\rho^{(2)}(x)$ is zero for $\|x\|<\eta^{-1/d}$.
In one-dimension, $Z^{(1)}(\alpha)=2\zeta(\alpha)$ (which leads to \eqref{eq:bound1} in the one-sided case).
For TDMA in 2 and 3 dimensions \cite{net:Zucker74},
\begin{IEEEeqnarray}{l} 
 \gamma^{(2)}=4\zeta(\alpha/2)\beta(\alpha/2)\theta \label{eq:gammad}\\
\gamma^{(3)}\approx(4\upsilon(\alpha) \zeta(\alpha/2-1/2)\beta(\alpha/2-1/2)
-\IEEEnonumber\\
\qquad\qquad 4\upsilon(\alpha)\zeta(\alpha-1)+8\zeta(\alpha/2)\beta(\alpha/2)-2\zeta(\alpha))\theta\IEEEnonumber
\label{eq:gammadb}
 \end{IEEEeqnarray}
where $\beta$ is the Dirichlet beta function and 
\[\chr{ \upsilon(\alpha)}=\frac{\sqrt{\pi}\Gamma(\alpha/2-1/2)}{\Gamma(\alpha/2)}.\]
In particular for $\alpha=4$,
\begin{align*}
\gamma^{(2)}&=\frac{2\pi^2}{3}G\theta \approx 6.03\,\theta\\
 \gamma^{(3)}&\approx \left(2\pi\zeta(3/2)\beta(3/2)
 				-2\pi A 
+\frac{4\pi^2}{3}G-\frac{\pi^4}{45}\right)\theta\\
& \approx 16.53\,\theta\,.
\end{align*}
where $G\approx0.916$ is Catalan's constant, and $A=\zeta(3)\approx1.202$ is  Ap\'ery's constant.
As expected, the spatial contention increases significantly from 2 to 3 dimensions.

\begin{figure}
\begin{center}
\includegraphics[width=.9\columnwidth]{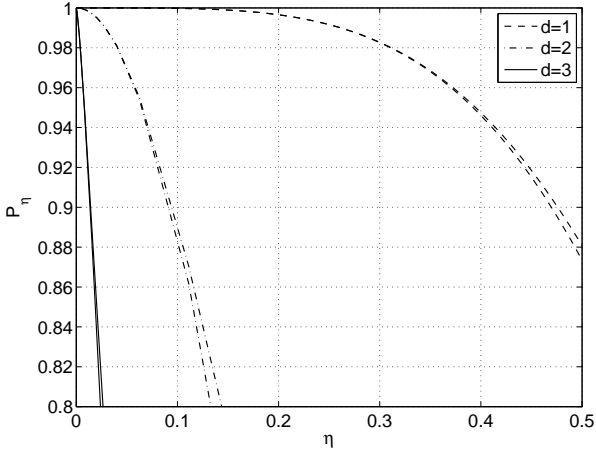}
\centering
\caption{Bounds \eqref{eq:boundd} for $m^d$-phase TDMA lattice network in $d$-dimensions for $\alpha=4$, $\theta=2$,
such that $1-\p_\eta=\Theta(\eta^{4/d})$.
The bounds are quite tight, and the decrease in the success probability for higher-dimensional networks is drastic.}
\label{fig:boundslattice}
\end{center}
\end{figure}

Results on $Z^{(d)}(\alpha)$ for other special cases are presented in \cite{net:Zucker74},
a general method to compute $Z^{(d)}(\alpha)$ efficiently can be found in \cite{net:Crandall98}.
In Figure \ref{fig:boundslattice}, the upper and lower bounds for the success probability are plotted for $d=1,2,3$ with $\alpha=4$, $\theta=2$.

\subsection{Extensions}
\label{sec:extensions}
\subsubsection{Different fading distributions}
\chr{The results in this paper can be easily extended  to any fading distribution between the typical receiver and the interferer's 
as long as the distribution of the received power from the intended transmitter $\S$ is exponential. In this case only the definition of $\Delta(x)$ has to be modified (generalized) to 
\[\Delta(x) = 1-\mathcal{L}_{\h{}}(\T\l(x)),\]
and the rest of the derivations remain the same.}
\ch{Generalizing the results to non-exponential $\S$ would require techniques that are significantly different from the ones used here.} 

\subsubsection{Swapping transmitter and receiver}
\begin{figure}
\centering
\includegraphics[width=.9\columnwidth]{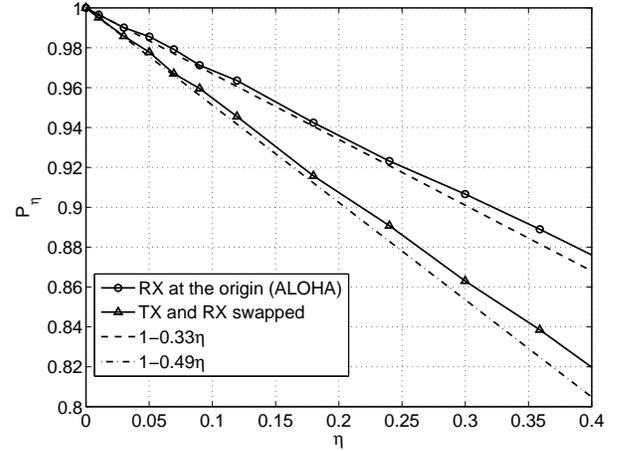}
\caption{\chr{Simulation of the outage probability of a hard-core point process with ALOHA, for a path loss law $\l(x)=\|x\|^{-4}$ and $\theta=2$. The hard-core radius is $h=1.5$, and the typical transmitter and the receiver are swapped.}}
\label{fig:ppp_ALOHA_swapped}
\end{figure}

\begin{figure}
\centering
\includegraphics[width=.9\columnwidth]{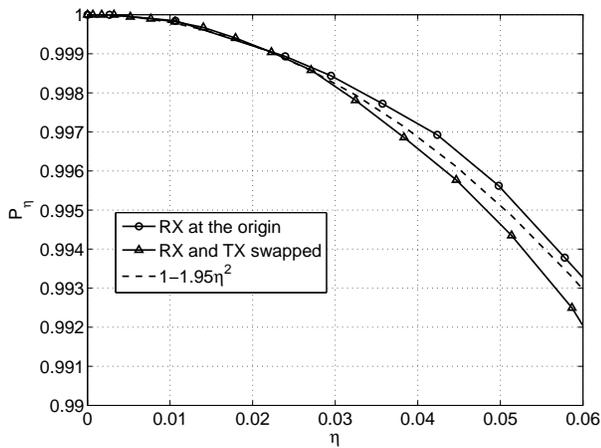}
\caption{\chr{Simulation of the success probability of a PPP of intensity $\lambda=0.3$
with CSMA, where the hard-core distance $h$ is adjusted to
achieve intensity $0.3\eta$, for a path loss law
$\l(x)=\|x\|^{-4}$ and $\theta=2$. The typical transmitter and the receiver at a distance $R=1$ are swapped,  and for comparison the asymptotic curve $1-1.95\eta^2$ is also shown. }}
\label{fig:ppp_csma_swapped}
\end{figure}

\ch{Until now, we have analyzed the case where we declared the typical node at the origin to be the receiver
under consideration. This way, the
notation was simplified, and there was no need to add an additional receiver node.}
 \chr{If instead, the typical transmitter is at the origin and its receiver is at a distance $R$, such that $\l(R)=1$, then 
\ch{the results change as follows:}
 \begin{enumerate}
 \item ALOHA MAC protocol (Theorem \ref{thm:aloha}): The new spatial contention parameter is
 \[\gamma_{\text{new}}=\lambda^{-1}\int_{\R^2} \rho^{(2)}(x)\Delta(x)\d x, \]
 where $ \Delta(x)= \frac{1}{1+\T^{-1}\l(x-y)^{-1}},$
and $y=(\tilde{R},0)$, where $\tilde{R}$ is the solution to $\l(\|x\|)=1$.
\item  Minimum-distance protocols (Theorem \ref{thm:two}):  The spatial contention parameter does not change and is given by (C.4).
 \end{enumerate}
 In Figure \ref{fig:ppp_ALOHA_swapped}, the success probability \ch{for ALOHA} is plotted for the Matern hard-core process with the transmitter and the receiver exchanged, and we can observe that the \ch{outage is still linear asymptotically}. \ch{For CSMA,} see Figure \ref{fig:ppp_csma_swapped} for an illustration of  $\p_\eta$  and the asymptotic curve for the CSMA Matern process, when the transmitter and the receiver are swapped.  As predicted, we observe that swapping the transmitter and the receiver has no effect on the asymptotic behavior.  }

\subsubsection{Varying the link distance}
\begin{figure}
\centering
\includegraphics[width=.9\columnwidth]{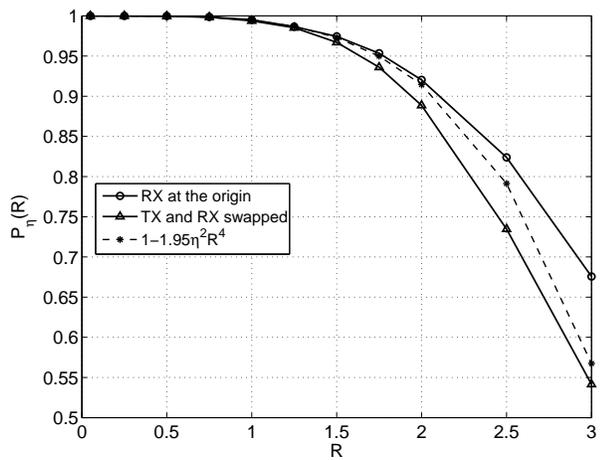}
\caption{\chr{Simulation of the success probability of Matern hard-core point process with CSMA  for varying TX-RX distance and for fixed $\eta$.   The tuning parameter $\eta$ is fixed at $\eta =0.052$ which corresponds to $h=4.5$ for $\lambda_p=1$. The path loss exponent is $4$ and $\theta=2$. }}
\label{fig:CSMA_R}
\end{figure}

\ch{Going back to the case where the typical {\em receiver} is at the origin,}
\chr{but its desired transmitter is at a general distance $R$, then \ch{we have the following changes:}
\begin{enumerate}
 \item ALOHA MAC protocol (Theorem \ref{thm:aloha}): The  spatial contention parameter is
 \[\gamma_{\text{new}}=\lambda^{-1}\int_{\R^2} \rho^{(2)}(x)\Delta(x)\d x, \]
 where 
 \[ \Delta(x)= \frac{1}{1+\T^{-1}\l(R)\l(x)^{-1}}\,.\]
\item  Minimum-distance protocols (Theorem \ref{thm:two}):  We have
\begin{equation}\gamma_{\text{new}}=\lim_{\eta \rightarrow 0} \T\l(R)^{-1} \lambda^{-1}\eta^{-\alpha/2-1} \int_{\R^2}
	  \l(x)\rho_\eta^{(2)}(x)\d x,
	  \label{eq:27}
	  \end{equation}
	  \ie, $\gamma$ gets multiplied by $\l(R)^{-1}$.
 \end{enumerate}
We observe that the spatial contention parameter $\gamma$ is scaled by $R^2$ in the case of  a PPP with ALOHA, while for
minimum-distance protocols (Matern hard-core processes) $\gamma$ gets scaled by $R^\alpha$. In Figure \ref{fig:CSMA_R}, the success probability  $\p_s$ of the CSMA protocol is plotted with respect to $R$ for $\alpha=4$, and the $R^4$ dependency \ch{is confirmed}.  For other node distributions and MAC schemes, we presume that $\gamma$ scales like $R^{2\kappa}$ (for large $R$), since scaling the space $\R^2$ by $R$ changes the density by $R^2$.
The interference scaling exponent $\kappa$ does not change by  swapping the typical transmitter and receiver, nor by changing the link distance.}

\begin{table}
\chr{
\begin{tabular}{|l|l|l|}
\hline
Node distribution& MAC& Tuning parameter: $\eta$ \\\hline
PPP &ALOHA&  Probability of transmission\\\hline
Hard-core process &ALOHA&Probability of transmission\\\hline
Hard-core process& CSMA&$\frac{1-\exp(-\lambda_p \pi h^2)}{\lambda_p \pi h^2}$\\\hline
PCP& ALOHA& Probability of transmission\\\hline
$d$-dim.  lattice&$m^d$ TDMA & $m^{-d}$\\\hline
\end{tabular}
\caption{This table summarizes the values of the tuning parameter $\eta$ for different initial node distributions and MAC protocols. See
Section \ref{sec:examples} for a detailed explanation of the point processes and MAC protocols under consideration.}
\label{tab:eta}}
\end{table}

\section{Conclusions}
\label{sec:conclusions}
We have derived the asymptotics of the  outage probability at low transmitter densities
for a wide range of  MAC protocols.  The asymptotic results are
of the form $\p_\eta\sim 1-\gamma \eta^\kappa$, $\eta \rightarrow 0$, where $\eta$ is the fraction
of nodes selected to transmit by the MAC. The two 
parameters $\kappa$ and $\gamma$ are  related to the network and to the MAC: $\gamma$
is the intrinsic {\it spatial contention} of the network introduced in
\cite{net:Haenggi09twc}, while $\kappa$ is the {\em interference scaling exponent} quantifying the
{\it coordination level} achieved by the MAC,
introduced in this paper. We studied $\p_\eta$ under the signal-to-interference ratio (SIR) model, 
for Rayleigh fading and with power-law path loss.

\ch{The numerical results indicate that for reasonable MAC schemes, the asymptotic result
approximates the true success probability quite well for $\p_\eta > 85\%$. In terms of $\eta$,
this means that the approximation is good as long as $\eta<\eta_{\max}$, where }
\[ \ch{\eta_{\max} \approx \left(\frac{0.15}{\gamma}\right)^{1/\kappa}} \,.\]
\ch{So for small $\gamma$ and $\kappa=\alpha/2$, the range of $\eta$ for which the approximation
is good is fairly large.}


Table \ref{table:reasonable} summarizes our findings and proposes a taxonomy
for reasonable MAC schemes. ALOHA belongs to class R1 ($\kappa=1$), irrespective of the underlying node
distribution; hard-core MACs such as reasonable TDMA and CSMA
are in class R3 ($\kappa=\alpha/2$); while soft-core MACs which guarantee a nearest-transmitter scaling
smaller than $\eta^{-1/2}$ are in class R2. Per Definition 1, the union of these three classes corresponds to
the set of reasonable MAC schemes.

\begin{table}
\centering
\begin{tabular}{|c|l|l|l|} 
\hline 
Class  & Range of $\kappa$ &  Scaling & Remark \\\hline
R1 & $\kappa=1$ & $\eta^{-1/2}$ on {\em average} & ALOHA \\\hline
R2 & $1\!<\!\kappa\!<\!\alpha/2$ & $\eta^{-a}$, $0<a<1/2$, a.s. & soft-core MAC\\\hline
R3 & $\kappa=\alpha/2$ & $\eta^{-1/2}$ a.s. & CSMA/TDMA\\\hline
\end{tabular}
\caption{Summary of the results and proposed MAC taxonomy for reasonable MAC schemes. The scaling
column indicates the nearest-neighbor distance scaling law as $\eta\to 0$.}
\label{table:reasonable}
\end{table}
%

Unreasonable MAC schemes are of less practical interest, but it is insightful to extend the taxonomy
to these MAC schemes and give the ranges of the parameters $\kappa$ and $\gamma$ that pertain to them.
The first class of unreasonable MAC schemes, denoted as U1, includes those MAC schemes for which the success
probability goes to $1$ but $\kappa<1$. 
This class is exemplified by the clustered MAC on a Poisson cluster process described in
Subsection \ref{sec:cluster2}.
Next, the example of highly clustered MACs (parent thinning) in Subsection \ref{sec:parent} shows that
 there exist MAC schemes for which
$\p_0=\lim_{\eta\to 0} \p_\eta< 1$. To incorporate such cases in our framework, we may generalize the
asymptotic success probability expression to
$\p_\eta \sim \p_0-\gamma\eta^{\kappa}$ for $\p_0\leq 1$. This is Class U2.
Lastly, there is an even more unreasonable class of MAC schemes, for which 
the success probability {\em decreases} as $\eta\to 0$, which implies that $\gamma<0$.
The TDMA scheme in \figref{fig:TDMA2} is an example of such an {\em extremely unreasonable} MAC scheme, which
constitute\ch{s} Class U3. A summary of this taxonomy of unreasonable MAC schemes is given in Table \ref{table:unreasonable}.


The different classes of MAC schemes can also be distinguished by the slope of the success probability at $\eta=0$.
A U1 MAC scheme has a slope of minus infinity, Class R1 has slope $-\gamma<0$, and Classes R2 and R3
have slope 0. Class U2 also has zero slope but $\p_0<1$, and U3 has a {\em positive} slope at $\eta=0$, which is
of course only possible if again $\p_0<1$.

Another way of looking at the different classes of MACs is their behavior in terms of repulsion or attraction of transmitters.
Class R1 (ALOHA) is neutral, it does not lead to repulsion or attraction, and whatever the underlying point process is,
as $\eta\to 0$, $\Phi_\eta$ approaches a PPP. Classes R2 and R3 induce repulsion between transmitters, which leads
to the improved scaling behavior. Classes U1 to U3, on the other hand, induce clustering of transmitters.

While it is possible to achieve $\kappa=\alpha/2$ (class R3)  for all point processes by choosing a good MAC scheme,
it is also possible to end up in classes U1, U2, U3 for all
point processes by choosing increasingly more unreasonable MACs.
This indicates that for $\kappa$ (and the sign of $\gamma$) only the MAC scheme matters, not the
properties of the underlying point process. As a consequence, {\em at high SIR, a good outage performance can
always be achieved, even if the point process exhibits strong clustering}. Conversely, if the MAC scheme is chosen
such that it favors transmissions by nearby nodes, the performance will be bad even if the points are arranged
in a lattice.

\begin{table}
\centering
\begin{tabular}{|c|l|l|l|l|} 
\hline 
Class & $\p_0 $ & Characteristic & Example \\\hline
U1 & $1$ & $0\!<\!\kappa\!<\!1$ & Cluster process in \figref{fig:clustered-b05} \\\hline
U2 & $<1$ & \ch{$\gamma\geq 0$} &  Cluster process with $b=0$  \\\hline 
U3 & $<1$ & $\gamma<0$ & TDMA scheme in \figref{fig:TDMA2} \\\hline
\end{tabular}
\caption{Extension of the MAC taxonomy to unreasonable MAC schemes. }
\label{table:unreasonable}
\end{table}

%


Our results also motivate
the following conjecture:
For all $0\leq \eta\leq 1$, we conjecture that the success probability $\p_\eta$ of any network with a
{\em reasonable MAC}  is bounded by
\begin{equation} \label{eq:conj}
1-\gamma \eta^\kappa\leq \p_\eta\leq \frac{1}{1+\gamma \eta^\kappa },
\end{equation}
where $\gamma>0$ and $\kappa\geq 1$ are two unique parameters.
The conjecture certainly holds in the case of ALOHA on the PPP (for all dimensions)
per \eqref{ps_ppp} and for (reasonable) TDMA on the lattice per \eqref{eq:boundd}.

\chr{When the transmitter and the receiver are swapped, this conjecture has to be modified to be valid for all $\eta$.
The conjecture as stated still holds for small $\eta$, though.}

\section*{Appendix}
\subsection{Proof of Theorem 1}
\begin{IEEEproof}
  Part 1 (lower bound): We first prove that $\kappa\geq 1$. We will show that 
$1-\p_\eta=O(\eta)$, which implies the result. 
From  \eqref{eq:succ_main} we have
\begin{eqnarray}
  \p_\eta &=& \E\exp\left(-\T\sum_{\x \in \G}\h{\x}\l(\x)\right)\\
   &\stackrel{(a)}{=}& \E\left[\prod_{\x \in
  \G}\mathbb{E} \exp(-\T\h{\x}\l(\x))\right]\nonumber\\
 &\stackrel{(a)}{=}& \E\left[\prod_{\x \in
  \G}\frac{1}{1+\T\l(\x)}\right]\\
 &=& \E\left[\prod_{\x \in  \G}1-  \Delta(\x)\right]\,,
\end{eqnarray}
where \chr{(a) is obtained by the independence of $\h{\x}$ and (b) follows from the Laplace transform of an exponentially
\ch{distributed} random variable, and} $\Delta(x)$ is given in \eqref{delta}.
Using the inequality 
\[\prod(1-y_i) \geq 1-\sum y_i,\quad y_i<1,\]
we obtain
\begin{eqnarray}
  \p_\eta &\geq& 1- \E\sum_{\x \in \G} \Delta(\x).
  \label{eq:proof12}
\end{eqnarray}
Hence
\begin{eqnarray}
  \frac{1-\p_\eta}{\eta^{1-\epsilon}} &\leq& \eta^{\epsilon-1} \E\sum_{\x
  \in \G} \Delta(\x)\nonumber\\
  &=&\eta^{\epsilon}\lambda\int_{\R^2}(\lambda\eta)^{-2}\rho_\eta^{(2)}(x)\Delta(x)\d
  x,
  \label{eq:proof13}
\end{eqnarray}
where $\rho^{(2)}_\eta(x)$ is the second-order product density of $\G$. 
\chr{Eqn.~\eqref{eq:proof13} follows from the definitions of the second-order product density and the 
 second-order reduced moment measure $\mathcal{K}_\eta(B)$.}
Tesselating the plane by unit squares $S_{kj} = [k,\  k+1]\times [j,\  j+1]$ yields
\[
  \int_{\R^2}\lt^{-2}\rho_\eta^{(2)}(x)\Delta(x)\d
  x = \lt^{-2}\sum_{(k,j)\in \Z^2}\int_{S_{kj}}\rho_\eta^{(2)}(x)\Delta(x)\d
  x\,. \]
Let $\Delta_{kj} \triangleq \Delta(\min\{\|x\|, x \in S_{kj}\})$. Since $\Delta(x)$ is a
decreasing function of  $\|x\|$, we have 
\begin{align*}
  \int_{\R^2}\lt^{-2}\rho_\eta^{(2)}(x)\Delta(x)\d x & < \lt^{-2}\sum_{(k,j)\in \Z^2}\Delta_{kj}\int_{S_{kj}}\rho_\eta^{(2)}(x)\d x\\
  &=\sum_{(k,j)\in \Z^2}\Delta_{kj}\mathcal{K}_\eta(S_{kj})\\
  &\stackrel{(a)}{<} C_{[0,1]^2}(\eta) \sum_{(k,j)\in \Z^2}\Delta_{kj}\\
  &\stackrel{(b)}{<} \infty,
\end{align*}
where $(a)$ follows from the transitive boundedness property of a PPD measure \ch{(see \eqref{eq:kbound} for
the definition of the constant $C_{[0,1]^2}(\eta)$),}
and  $(b)$ follows from Condition \eqref{conda} and since for $\alpha>2$, $ \sum_{(k,j)\in \Z^2}\Delta_{kj} <\infty$.
Hence it follows from \eqref{eq:proof13}  that 
\[\lim_{\eta \rightarrow 0}  \frac{1-\p_\eta}{\eta^{1-\epsilon}}  =0 \,,\]
which concludes the proof of the lower bound.\\
Part 2 (upper bound): Next we prove that $\kappa \leq \alpha/2$. We will show that 
$1-\p_\eta=\Omega(\eta^{\alpha/2})$,
which implies the result. 
The success probability is 
\begin{eqnarray*}
\p_\eta &=& \E\left[\prod_{\x \in
  \G}\frac{1}{1+\T\l(\x)}\right]\\
  &\leq& \E\left[\prod_{\x \in
  \G\cap B(o,R\eta^{-1/2})}\frac{1}{1+\T\l(\x)}\right]\\
  &\leq&\E\left[\frac{1}{1+\T\l(R \eta^{-1/2})}\right]^{\G(B(o,R\eta^{-1/2}))}
\end{eqnarray*}
As $\eta \rightarrow 0$, $\l(R \eta^{-1/2}) \sim
R^{-\alpha}\eta^{\alpha/2}$, and using the identity $(1+x)^{-k}\sim 1-kx$
for small $x$ we obtain
\begin{eqnarray*}
  \lim_{\eta\rightarrow 0}\frac{1-\p_\eta}{\eta^{\alpha/2}} &\geq&
  \lim_{\eta\rightarrow 0}\E[\G(B(o,R\eta^{-1/2}))]
 \T R^{-\alpha}\\
&=&\lim_{\eta\rightarrow 0}\eta \lambda K_\eta(R\eta^{-1/2})\T R^{-\alpha}\\
 &\stackrel{(a)}{>}& 0,
\end{eqnarray*}
where $(a)$ follows from \eqref{condb}. This concludes the proof of the
upper bound on $\kappa$.
\end{IEEEproof}

\subsection{Proof of Theorem 3}
\begin{IEEEproof}
  From \eqref{eq:main_err} we have
	\[\p_\eta=\E\prod_{\x\in\G}\frac{1}{1+\theta\l(\x)}.\]
	It is easy to observe that 
	\begin{equation}
 \EP\exp\Big(-\T\sum_{\x \in \G}\l(\x)\Big) \leq \p_\eta\leq
	\EP\Big[	\frac{1}{1+\T\sum_{\x \in \G}\l(\x)} \Big] \,.
	\label{eq:inequality}
  \end{equation}
Let $\I_\eta=\sum_{\x \in \G}\l(x)$.
	By Jensen's inequality we have 
	\begin{align*}
	\p_\eta & \geq \exp\left(-\T\,\E\I_\eta\right)  \\
	  &= \exp\left(-\T\eta^{-1}\lambda^{-1}\int_{\R^2}
	\rho^{(2)}_\eta(x)\l(x)\d x\right)\,.
	\end{align*}
	By Condition \eqref{cond31}, we obtain
	\[\lim_{\eta\rightarrow 0}\frac{1-\p_\eta}{\eta^{\alpha/2}} \leq \gamma.	\]
	To show the converse, we upper bound $\p_\eta$ as  
	\begin{eqnarray}
	  \p_\eta&\leq&\E\left[\frac{1}{1+\T\I_\eta}\right]\nonumber\\
	   &\leq&1-\E[\T \I_\eta] + \E[\T^2\I_\eta^2]\,.
	  \label{eq:455}
	\end{eqnarray}
	It is easy to show that 
	\begin{align}
	  \lambda \eta \E[\I^2_\eta]
	  &=\int_{\R^2}\l^2(y)\rho^{(2)}_\eta(y)\d
	  y\nonumber\\
	  &+\int_{\R^2}\int_{\R^2} \l( x) \l( y)  
	\rho^{(3)}(x,y)\d x\d y,
	\end{align}
	which, combined with Conditions \eqref{cond32} and \eqref{cond33}  implies $\E[\I^2_\eta]\eta^{-\alpha/2}\rightarrow 0$ as
	$\eta\rightarrow 0$.
	Hence it follows from \eqref{eq:455} that 
	\[\lim_{\eta\rightarrow 0}\frac{1-\p_\eta}{\eta^{\alpha/2}} \geq \lim_{\eta\rightarrow 0}\T
	\eta^{-\alpha/2}\E[\I_\eta] = \gamma.\]
\end{IEEEproof}

\subsection{Proof of Corollary 1}
\begin{IEEEproof}
We show that in this case, Condition \eqref{cond31} holds.
We focus on the singular path loss law $\l(x)=\|x\|^{-\alpha}$; the
other cases follow in a similar manner. Let $C_1\triangleq \gamma\lambda/\theta$.
We have
\[C_1 = \int_{\R^2} \|y\|^{-\alpha}\rho_\eta^{(2)}(y\eta^{-1/2})\eta^{-2}\d y.\]
Since the support of $\rho^{(2)}_\eta(x)$ lies in $B(o,c\eta^{-1/2})^c\triangleq\R^2\setminus B(o,c\eta^{-1/2})$ we have
\[C_1 = \int_{B(o,c\eta^{-1/2})^c} \|y\|^{-\alpha}\rho_\eta^{(2)}(y\eta^{-1/2})\eta^{-2}\d y.\]
Using the substitution $y\eta^{-1/2} \rightarrow x$ we obtain
\[C_1 = \eta^{-1-\alpha/2}\int_{B(o,c\eta^{-1/2})^c}\|x\|^{-\alpha}\rho_\eta^{(2)}(x)\d
x.\]
Letting $A_k =B(o,ck\eta^{-1/2})$, we have
\begin{align*}
  C_1 &=\eta^{-1-\alpha/2}\sum_{m=1}^\infty \int_{A_{m+1}\setminus A_m}\|x\|^{-\alpha}\rho_\eta^{(2)}(x)\d
x \\
&\leq\eta^{-1-\alpha/2}\sum_{m=1}^\infty (cm\eta^{-1/2})^{-\alpha}\int_{A_{m+1}\setminus A_m}\rho_\eta^{(2)}(x)\d
x\\
&\stackrel{(a)}{=}\eta\sum_{m=1}^\infty
(cm)^{-\alpha}[K_\eta(\eta^{-1/2}(m+1))-K_\eta(\eta^{-1/2}m)],
\end{align*}
where $(a)$ follows from the identity $\int_{B(o,R)}\rho_\eta^{(2)}(x)\d x
\equiv\lt^2 K_\eta(R)$.
For large $R$, we have $K(R)\sim \pi R^2$, hence 
\[K_\eta(\eta^{-1/2}(m+1))-K_\eta(\eta^{-1/2}m)\sim \pi\eta^{-1} (2m+1),\]
and thus $C_1<\infty$ for $\alpha>2$. Using a similar method, Condition \eqref{cond32} 
	 can also be shown to hold in this case. So in CSMA networks
	whose inhibition radius scales as $\eta^{-1/2}$, the conditions in Theorem
	\ref{thm:two} are satisfied and $\kappa=\alpha/2$. 
\end{IEEEproof}

\bibliographystyle{IEEEtran}


\begin{IEEEbiography}
[{\includegraphics[width=1in,height=1.25in,clip,keepaspectratio]{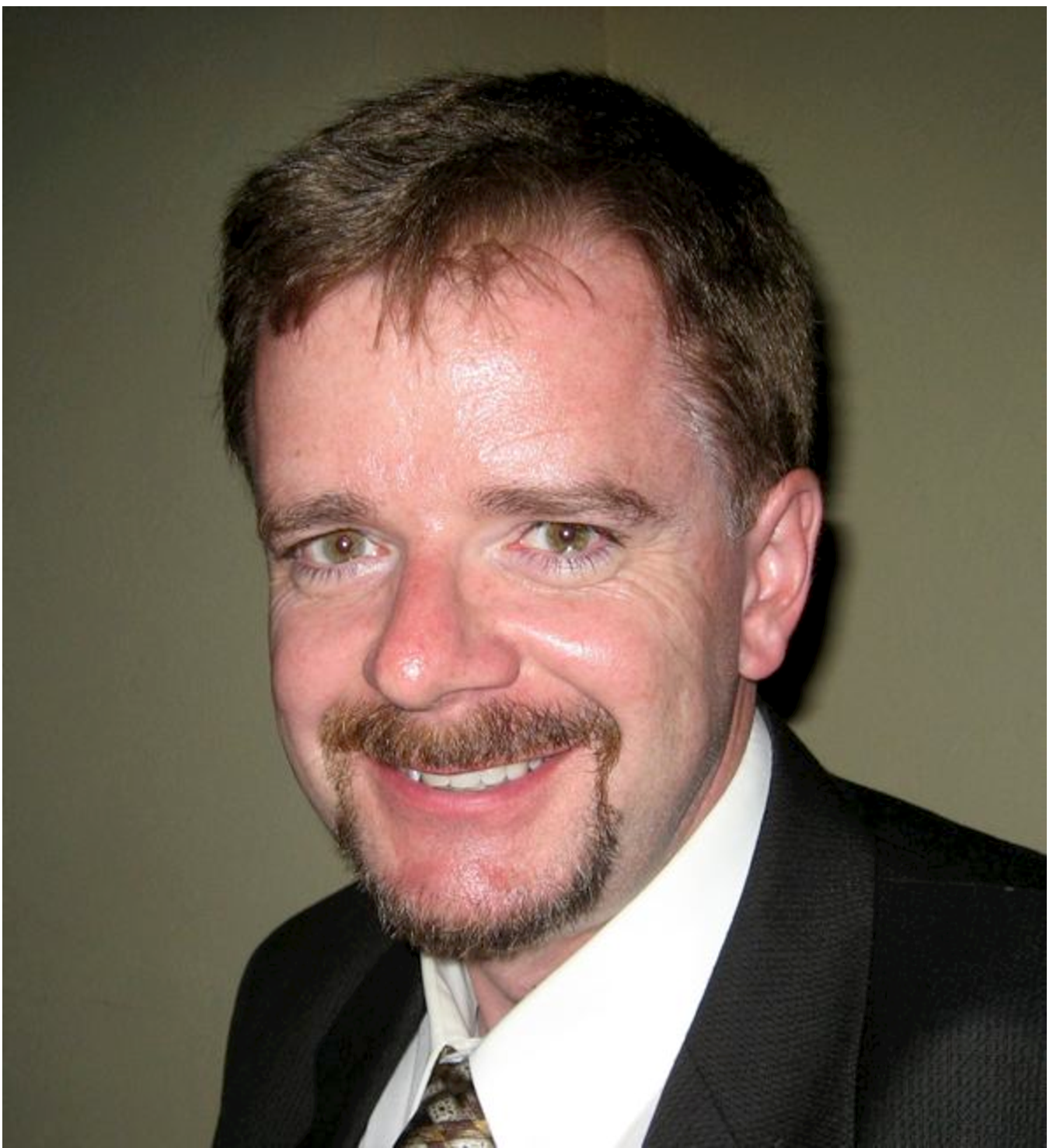}}]
{Martin Haenggi}
(S'95, M'99, SM'04) is an Associate Professor of Electrical Engineering at
the University of Notre Dame, Indiana, USA. He received the
Dipl. Ing. (M.Sc.)  and Ph.D. degrees in electrical engineering from
the Swiss Federal Institute of Technology in Zurich (ETHZ) in 1995 and
1999, respectively.  After a postdoctoral year at the Electronics
Research Laboratory at the University of California in Berkeley, he
joined the University of Notre Dame in 2001.  In 2007-08, he spent a Sabbatical Year
at the University of California at San Diego (UCSD). For both his M.Sc. and his
Ph.D. theses, he was awarded the ETH medal, and he received a CAREER
award from the U.S.~National Science Foundation in 2005 and the 2010
IEEE Communications Society Best Tutorial Paper award. He served as a member
of the Editorial Board of the Elsevier Journal of Ad Hoc Networks from 2005-08, as a
Distinguished Lecturer for the IEEE Circuits and Systems Society in 2005-06, and as a Guest Editor
for the IEEE Journal on Selected Areas in Communications in 2009. Presently he is an Associate
Editor the IEEE Transactions on Mobile Computing and the ACM Transactions on Sensor Networks.
He is a co-author of the monograph Interference in Large Wireless Networks (NOW Publishers, 2008) \cite{net:Haenggi08now}.
His scientific interests include networking and wireless
communications, with an emphasis on ad hoc, sensor, mesh, and cognitive networks.
\end{IEEEbiography}
\begin{IEEEbiography} [{\includegraphics[width=1in,height=1.25in,clip,keepaspectratio]{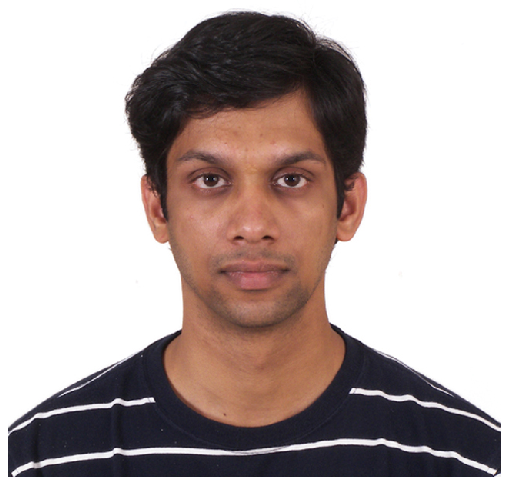}}]{Radha Krishna Ganti}
(S'01, M'10) is a Postdoctoral researcher in the Wireless Networking and Communications Group at UT
Austin. He received his B. Tech. and M. Tech. in EE from the Indian Institute of Technology, Madras, and a Masters in
Applied Mathematics and a Ph.D. in EE from the University of Notre Dame in 2009. His doctoral work focused on the spatial
analysis of interference networks using tools from stochastic geometry. He is a co-author of the monograph
Interference in Large Wireless Networks (NOW Publishers, 2008) \cite{net:Haenggi08now}.
\end{IEEEbiography}
\begin{IEEEbiography} [{\includegraphics[width=1in,height=1.25in,clip,keepaspectratio]{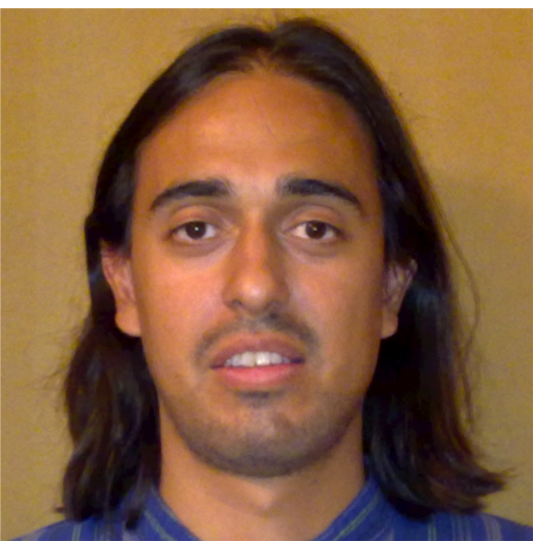}}]{Riccardo Giacomelli}
is a Post-Doctoral researcher at the Politecnico di Torino, Turin, Italy.
He received his M.Sc.~and  Ph.D.~degrees from the Politecnico di Torino in 2006 and 2010, respectively.
In 2009, he spent 6 months as a visiting Ph.D.~student at the University of Notre Dame, Indiana, USA.
His research focus is on the analysis and design of wireless networks.
\end{IEEEbiography}

%

\end{document}